# SYSTEMATIC ATTACK SURFACE REDUCTION FOR DEPLOYED SENTIMENT ANALYSIS MODELS

Josh Kalin[1], David Noever[2], and Gerry Dozier[1]

[1]Department of Computer Science and Software Engineering, Auburn University, Auburn, AL, USA
jzk0098@auburn.edu, doziegv@auburn.edu
[2]PeopleTec, Inc, Huntsville, AL, USA
David.Noever@peopletec.com

## ABSTRACT

*This work proposes a structured approach to baselining a model, identifying attack vectors, and securing the machine learning models after deployment. This method for securing each model post deployment is called the BAD (Build, Attack, and Defend) Architecture. Two implementations of the BAD architecture are evaluated to quantify the adversarial life cycle for a black box Sentiment Analysis system. As a challenging diagnostic, the Jigsaw Toxic Bias dataset is selected as the baseline in our performance tool. Each implementation of the architecture will build a baseline performance report, attack a common weakness, and defend the incoming attack. As an important note: each attack surface demonstrated in this work is detectable and preventable. The goal is to demonstrate a viable methodology for securing a machine learning model in a production setting.*

## 1. INTRODUCTION

This paper is structured into six separate sections: Introduction, Background, Approach, Evaluation, Future Contributions, and Conclusions.

Sentiment Analysis (SA) [1] is the task of analyzing text to provide a classification such as positive, negative, or neutral for a given sample. SA is subdivided further into categories such as polarity, subject, and toxicity. Companies and organizations use these technologies to moderate their websites, apps, and comment sections [2]. Adversarial attacks, in the context of this work, refer to any input that allows an adversarial actor to trick a classification system. Modern SA Systems use machine learning (ML) and are susceptible to adversarial attacks [3]. Recently, the Natural Language Processing (NLP) community has explored how to create models that can handle bias in training data; for instance, content-aware models are an example of a system that can interpret bias in the data and correctly classify sentiment [4]. Given the challenging nature of SA with this type of data, the goal is to demonstrate a simple and repeatable process for creating a model baseline, attacking the model, and defending against the incoming attacks.

Toxicity Classification [5] is a SA technique to understand the malicious intent of text based on words and content in the message. These SA techniques use ML and Deep Learning (DL) to classify the toxicity or polarity of a tweet [6]. The first SA technique used in this paper is the Sentiment140 SA API [7]. Sentiment140 originated as a paper from the early 2010s and was later developed into an API by a Stanford Team [8]. Perspective, the second SA API used, is built and maintained by Google's Jigsaw team [9]. The Perspective API is a black box ML model that relies on a transformer and other deep learning technologies to classify sentiment. The Perspective API focuses on toxicity analysis for social media-based comments [10]. Each SA API uses machine learning to provide sentiment classification. We have no connection or insight into the underlying models other than published papers or websites. Further, there are limitations on the number of queries per second and per day.

### 1.1. Challenges

The dataset used in this work creates a unique challenge. The Conversation AI Team, funded in conjunction with Jigsaw and Google, created a dataset around toxicity, biases, and threats in comment sections [11]. The JigSaw Toxic Bias dataset is a set of publicly released comments augmented with new labels for ML tasks. It has a wide range of different toxicity classifications such as severe toxicity, obscene, identity attack, insult, and threat. In the last year, the Conversation AI team has augmented JigSaw with additional categories including gender, sexual orientation, and religious identity [9]. The new evaluation categories were added to combat inherent biases that are included in the data but do not represent a negative sentiment. This dataset is part of a challenge on the Kaggle competition page for creating the best classification models around toxicity. The top-scoring models used ensembles of DL models to get the highest classification scores [12]. Since the newest classification techniques for this dataset use ML, they are susceptible to adversarial methods [13]. Adversarial Methods demonstrated in the Evaluation section are focused on discovering attacks that negatively affect the classification capability of the underlying system. This paper will focus on the challenge of evaluating the attack surface of a single attack vector and defending the model from this incoming attack.

### 1.2. Contributions

Single Character attacks vectors are a direct analog to single-pixel attacks in the image domain - for instance, single-pixel attacks have demonstrated effects on classification, reinforcement learning, and other state-of-the-art image technologies [14-15]. We demonstrate the efficacy of single character attacks (1 or many) on these sentiment text classifiers and how to protect the underlying system. These simple attacks can reduce the ability of systems to filter and curate online media platforms. This paper focuses on demonstrating this new architecture to build a baseline of the performance of each API, attack the models with single character substitution/insertion attacks in the text domain, and provide a defense plan for these attacks. The remainder of this paper is as follows: Section 2 presents the related work, Section 3 develops the approach, Section 4 discusses solutions, and Section 5 closes with Conclusions and Future Work.

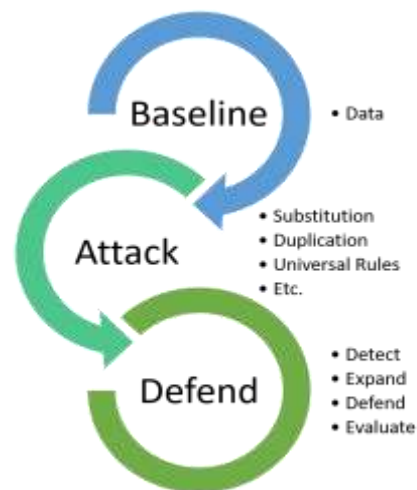

Figure 1: BAD System: Baseline, Attack, Defend for Protecting Machine Learning Models

## 2. RELATED WORK

There are three key background areas: Build, Defend, Attack Systems from the Cyber Security Domain, the Sentiment Analysis, and black-box models.

## 2.1. Build, Defend, Attack Systems

Build, Attack, Defend is a construct built around breaking down Cyber Security problems into actionable problem spaces for teams to digest [16]. There are relevant team constructs around Color teams for attacking, defending, and designing systems:

- Red Team: Ethical hacking of a target system

- Blue Team: A group of people building defenses against the attacks on the model

- Green Team: A blend of the red and blue teams where the group will simultaneously run both attacks and create defenses for those attacks [17].

The focus is on applying the green team construct to improve ML model development. By utilizing adversarial design knowledge (Red Team) and model building knowledge (Blue Team), the Green Team can propose defenses to the underlying model designs or production pipelines that secure the model from outside attacks.

## 2.2. Sentiment Analysis

The field of SA has progressed rapidly due to the expansion of social media platforms. Among internet moderation applications, SA is a required piece of policing social media and comment sections due to the large volume of comments on these sites [18]. With state-of-the-art SA systems improving, it became evident that there is an unintended bias built into the dataset of comments and tweets [10]. Overfitting to words in bias dataset has led to embarrassing results for production SA. The famous example with Perspective API is "I am a gay black woman" which carried a 95% toxicity in 2017 and still contains a toxicity score of 44% as of writing this paper [19]. State-of-the-art SA papers are dominated by ensembles of transformer technologies for this particular problem set [20]. As a note: Each of the SA classifiers inherits weaknesses of the underlying language models used for the classification tasks.

## 2.3. Black Box Models

A black box model in ML is any model that an end-user only has access to inputs and outputs [21]. Each ML black box model in this work allows an end-user to interact with it through JSON inputs and outputs. There are also limitations to the number of queries per user per system. Two black box SA systems were selected in this paper: Sentiment140 and Perspective API. Sentiment140 is based on a technical report which collected 1.6million tweets to survey ML techniques in the SA domain in 2008 [8]. The Sentiment140 team has maintained the API as a historical benchmark for future SA systems but provides no explicit details on the exact implementation of the API. The Sentiment140 team implemented the paper. In contrast, the Perspective team provides toxicity scores for multiple categories through their API. With Perspective, an end-user can request classification probability scores for each class and can therefore evaluate the efficacy of each attack. The Perspective Team does not provide details on their machine learning models.

## 3. APPROACH

There are numerous areas of modeling where an adversarial actor can attack. For simplicity, the focus is on inference-based attacks. Attacks on the inference pipeline exploit weaknesses of data used for training and learned weights of the model. For example, there are simple attacks like substitution, replication, and insertion that easily fool current classification models. A recent paper proved universal rules for fooling text-based classification systems are effective for multiple tasks in NLP [22]. The Evaluation section demonstrates a BAD architecture focused on an inference-based attack for each API. Figure 1 shows the general flow of implementing the BAD architecture for an inference attack surface of a SA Model. The following sections discuss each of the Build, Attack, and Defend core components in detail.

### 3.1. Holistic Approach: Introducing BAD Architecture

Every machine learning team wants to understand the model's vulnerability to adversarial attacks. The BAD Architecture proposes three key steps. First, a team needs to understand the baseline performance of the model by asking questions like the following:

- How does the model act with regular and irregular data?

- Are there known weaknesses or limitations?

- Are those limitations and risks mitigated?

Next, a team needs to understand the impact of each attack by exercising each vulnerability in the model. Last, after understanding the baseline performance and attacking their model, the team will need to propose and implement those defenses to protect their production process. In practice, this entire architecture is repeatable and expandable depending on the scope of the team.

### 3.2. BUILD a baseline of our target system

A core componentof a ML production system is to understand performance under normal conditions. With the BAD Architecture, each team should also note the known limitations of the model. For instance, some systems do not inherently return real scores for words not in the original training data (example: Word2Vec) [23]. A team must be upfront and understand the impact of design decisions on how a ML system has been designed. To baseline a ML system, it is also important to experiment with data that the system is expected to operate on regularly. If possible, it is also expected to document any edge cases that would be hard for the system to classify. Using an SA black-box model with the JigSaw dataset is a perfect example baseline case for the Build, Attack, Defend Architecture. The data contains toxic edge cases where it is hard to judge the intent of the underlying message. The advantages of creating a systematic baseline are shown in the Evaluation section when edge cases are exploited.

### 3.3. ATTACK System weaknesses and inefficiencies

Each API has a public page and allows anyone to sign up for basic services. Even with basic access, it is possible to circumvent these systems with a limited number of queries and the Python programming language [24]. Adversarial Character Attacks in the NLP field revolve around changing one or more characters while maintaining the original intent to a human annotator. An Adversarial Attacks using character attacks attempt to direct the decision boundary of the underlying detector in a way that is beneficial to the attacker [25]. For a SA system, this would use substitution attacks to avoid the detection of negative or toxic comments in a social media environment. If bad actors understood how to substitute common character and reduce their toxicity, then it becomes easy for them to use hate speech (as an example). There are two areas in the character attack space applied here: substitution and duplication. The example Attack system demonstrates simple substitution attacks:

1) Create a dictionary that contains vowel to alpha-numeric (for instance e:3)

2) For every vowel in the sentence, replace a single instance from the string and store in an array

3) For every string in the array, evaluate sentiment through public-facing API

4) Evaluate the number of times a single character changed the score or decision made by the black-box model

And, for the duplication Attack, the same process is replicated with only a change in the attack vector:

1) Create a dictionary that contains vowels to duplicated vowels (for instance e:ee)

2) For every vowel in the sentence, replace a single instance from the string and store in an array

3) For every string in the array, evaluate sentiment through public-facing API

4) Evaluate the number of times a single character changed the score or decision made by the black-box model

Given the nature of black-box models, each API only provides the probability of toxicity or polarity without additional feedback. With Sentiment140, we are provided three states of polarity: negative, neutral, and positive. There are no percentages of each classification; rather the API simply provides the highest binary classification value. It is only possible to show if a classification can move from one category to another. With Perspective, the actual probability of each classification category is available for each request. Therefore, it is possible to see the decrease or increase in confidence for a given input. The Evaluation section shows the baseline and delta results for each of the attacks.

### 3.4. Hello World of the BAD Architecture

Figure 2 shows how the system will operate on the simplest incoming toxic phrase. In this example, the 'I hate people' example demonstrates the Build, Attack, and Defend pipeline. Applying the Perspective API, this string scores an 82% toxicity and negative score on the Sentiment140 system. When a Red Team attacks the model with a single character substitution attack of "a" to "@", the toxicity of the comment goes down to 30%. The Green Team's goal is to break apart each attack vector and create a more robust system against adversarial attacks. The Defend section will cover possible strategies for combating simple attacks.

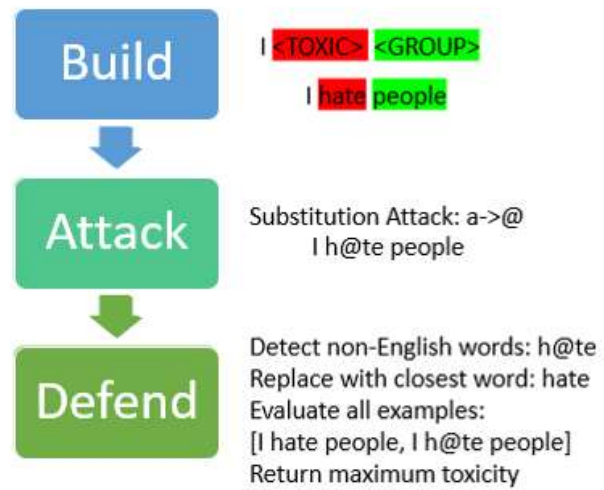

### 3.5. DEFEND System from Targeted Attack

The Green Team will summarize the baseline and adversarial results to create a defense plan. Typically, there are simple ways to mitigate attacks. For example, a back-end developer can create limitations around what types of requests can be made. Take the commentary systems on a forum: if they use the Perspective API, it would be straightforward to add a few rules to reduce the ability of an attacker to use substitution attacks (Evaluation section covers a basic implementation). When production systems are used as 'download and deploy', creates a wide variety of pointed and effective attacks vectors for adversarial actors. In practice, there are the following crucial steps:

1) Detect: Detect and catch the adversarial text

2) Expand: Expanding the text to include the possible meanings of the originator

3) Defend: Process each result and store for future analysis

4) Evaluate: Check each result and return tune how the team wants the system to respond to attacks

This process cannot stay stationary. Bad Actors are constantly working to find new and inventive ways to break ML systems. The goal of this process is to create development architecture that can be deployed ML model development. In our Evaluation section, we focus on SA and the way we use this system to evaluate the Sentiment140 and Perspective API systems.

## 4. EVALUATION

Each API provides the ability to send one query per second (1 QPS). There are limitations to the number of adversarial examples we could present to the Perspective API for instance which had a limit of 100 one-second queries. In this instance, a local model is trained and a model is attacked. Then, the potent attacks that fooled the local SA system are used against the black-box model. In practice, adversarial examples were drawn from the training set as a sample of the one hundred top toxic examples for each category of JigSaw. There is a section for Sentiment140 and Perspective where the Build, Attack, Defend Architecture is explained in detail.

### 4.1. Sentiment140

The Sentiment140 API has a large limit to queries (approximately 5000 items per query per second). There is a maximum limit of around 800,000 scored queries in a given time period (experimentally derived). Experiments are limited to a few permutations of substitution and insertion attacks per input row. In the Build section, the baseline results of the Sentiment140 model with the Jigsaw dataset are covered. In the Attack section, experimental results with simply applying substitution and insertion attacks are explored. The baseline experiments for the JigSaw dataset will be paralleled between the two systems - choose one hundred samples from each toxic category with a 50% or above toxicity and then attack the classification of each toxic row. For Sentiment140, we are given binary results for each experiment. Every returned row provides a category of positive, negative, or neutral. The baseline results are seen in Figure 3 in the Build Section. The classification binary score for these toxic comments is the majority in the neutral and positive categories. For example, 69 percent of the threat category is classified as either neutral or positive. As a note, each toxic input has been annotated by a human to include the label. Sentiment140 does still miss out on a large chunk of the proper classifications of negative for each one of these input rows. The next experiment will show how substitution attacks push the decision boundaries for this model in a different direction.

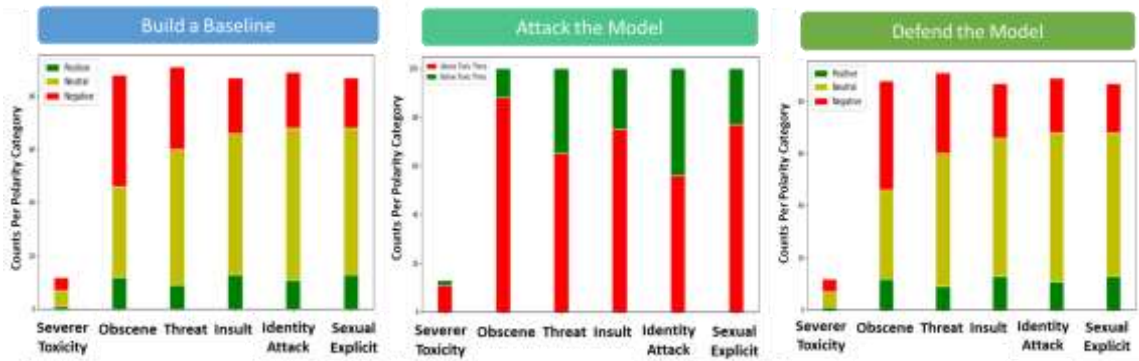

Figure 3: Sentimet140 Performance Summary using the Build, Attack, Defend Development Architecture

### 4.1.2. ATTACK: Results with Substitution Attacks

The substitution attacks had an interesting effect. The neutral classifications almost universally transformed into negative sentiment. The threat category, for instance, went from 31 percent negative to a 69 percent negative for the one hundred samples included here. In Figure 3, the massive change in the polarity is evident in each category. If the goal was to simply push all polarities to positive, then additional experiments with additional techniques would need to be explored. This work, demonstrates two things:

1) The Sentiment140 algorithm, with this open API, is not well equipped to deal with the bias inherent in modern social media commentary.

2) The decision boundary between neutral and negative is much closer than anticipated with simple substitutions changing neutral polarity into negative polarity.

Sentiment140 was never meant to work with social commentary with this level of toxic bias. Since these are black-box models, there is no opportunity to improve the performance of the underlying system. This highlights the core advantage of applying this architecture. With simple access, the underlying ML model can be evaluated and tested. For defense, both APIs are covered under the Adversarial Attack Surface Reduction section.

### 4.2. Perspective API

The Perspective API allows one text field per query per second. There is a limit to the number of daily queries but it was not a problem in the experiments. The first step is to create a baseline performance on 100 examples from each of the toxicity categories. Then, attack the same sample of 100 with character attacks (alpha-numeric and duplication) in each category to understand how much degradation can be introduced with simple character attacks.

### 4.2.1. BUILD: Baseline JigSaw Performance with Perspective

There are two separate experiments run during the baseline stage. A baseline of production models against the Jigsaw data is evaluated. This data is pulled as a sample, straight from the training set, with each toxic category measuring at 50% toxicity or greater for that category (same process as with Sentiment140). If the model were able to classify them appropriately, every one of the examples we pulled would be toxic (red). Our results, shown in Figure 4, show that there is still work to be done in terms of getting full coverage of even just the hundred selected training examples.

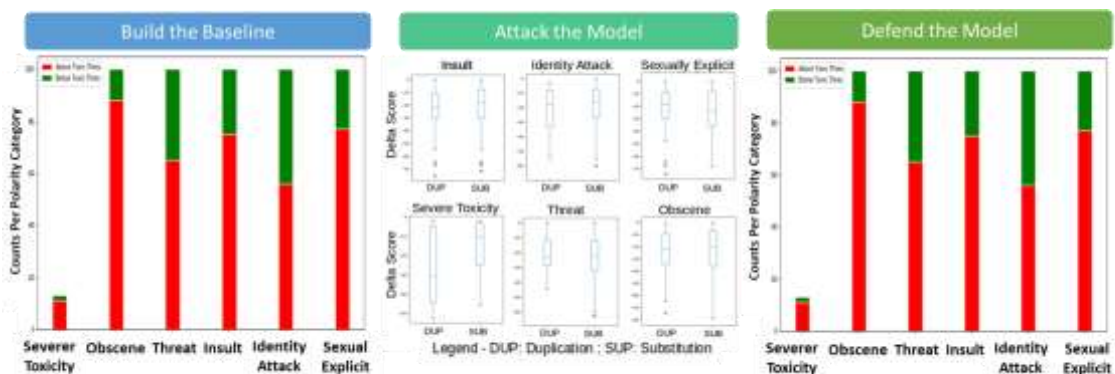

Figure 4: Perspective API Performance Summary using the Build, Attack, Defend Development Architecture

### 4.2.2. ATTACK: Results with Substitution Attacks

First, for every vowel, substitution and duplication attacks at each vowel position are applied. The Perspective API is robust to these types of simpler attacks, as the delta between the original score and the attacked score was less than 5% in the experiments. Only a single example went below 4% delta in its delta score. This portion of the attack space will need more exploration. One can think about the single character changed once in a string as a single-pixel attack. Only a single pixel is being changed throughout the entire picture. There is another approach to the single character problem - in this attack, consider changing the same character and replace or duplicate it at every instance. The analog to this attack would be changing a certain color pixel throughout the entire image. This exercise is left to another time. The final experiment, highlighted in Figure 4, explores changing vowels to an alpha-numeric or simply duplicating them. Substitution and Duplication attacks are discussed throughout the literature as rudimentary but effective tools when surveying the adversarial surface of these models [22]. In this experiment, the focus was on vowel substitution and duplication attacks on the 100 samples on a per-category basis. In every case, there was at least one example of a -70% reduction in the toxicity score, effectively taking the sentiment from toxic to non-toxic. By using the Attack Surface Reduction steps for our Defend step, it is possible to completely negate the original attacks demonstrated in the last two sections.

### 4.3. ATTACK SURFACE REDUCTION FOR BOTH APIs

For each of the substitution attacks, there are simple code changes offered for filtering each of these results. In fact, by utilizing these filtering techniques, it is possible to restore the original classification accuracy of the system. Unfortunately, this work does not focus on improving the black box models. The goal is to demonstrate vulnerabilities inherent in these systems and propose an architecture for production systems to protect the efficiency of their systems.

### 4.3.1. Attack Surface Reduction: Substitution Attacks

This work features two specific types of character attacks - replacement and duplication. The focus is limited to English in this effort although these methods should translate to other languages. First, for detection, the user can detect all non-English words. Multiple models will provide the nearest word or words in a corpus of available words. A simple Defend preprocessing pipeline before inference would be as follows:

1) Find all non-English words by using standard NLP libraries such as NLTK [26]

2) For each non-English word, find nearest neighbor words using algorithms such as Word2Vec [27]

3) Create an array of text with the non-English words replaced with the top N candidates

4) Evaluate the array of text and take the max or min score for classification

In practice, there are commonly misspelled words that can be safely ignored. Using this Defend pipeline can increase the number of candidates for inference but provides robustness to the attacks shown in this paper. As another method, the systems can maintain a set of common substitutions such as alpha-numeric substitutions using alpha-numeric characters or other simple dictionary lookups. Each detected "Attack" should be stored and evaluated by the Green Team periodically to ensure that the pipeline is working as designed.

### 4.4. Limitations

In each example, the attacks were simple character to character mappings. Zero-Day attacks in the cyber realm refer to attacks that are not yet protected against and allow a hacker unfettered access to a system. In the machine learning realm, there are adversarial 'zero-day' attacks that are manipulating the output of the model. These Zero-Day attacks are difficult to anticipate and protect against in practice. This architecture currently relies on known attacks on models for protections and does not actively search an adversarial surface for the model.

## 5. CONCLUSIONS AND FUTURE WORK

This work demonstrates that deployed sentiment models are susceptible to simple substitution attacks on single characters and can be effectively defended from each substitution attack using the BAD architecture. Because these substitutions are simple character to character mappings, they are mitigated by detecting non-English words, creating candidates for sentiment analysis, and taking the maximum toxicity in our examples. Further work in this area will focus on looking at model attacks like weight poisoning attacks on classification systems.

Weight Poisoning Attacks on Pre-trained Models [28] is a recent paper that uses vulnerabilities in pre-trained models and strikes me as dangerous to all black box models that are not actively defending against those types of tasks. A future direction could be to develop a data augmentation method or model structure that makes weight poisoning attacks reduces the efficacy of weight poisoning attacks, During the defend phase, automated methods for detecting and correcting poisoned words could use transformer models to find and propose corrected word.

### ACKNOWLEDGMENTS

Special thanks to the medical folks out there during the COVID-19 pandemic

**Authors**

**Josh Kalin** is a physicist and data scientist focused on the intersections of robotics, data science, and machine learning. Josh holds degrees in Physics, Mechanical Engineering, and Computer Science.

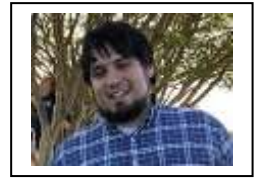

**David Noever** has research experience with NASA and the Department of Defense in machine learning and data mining. He received his Ph.D. from Oxford University, as a Rhodes Scholar, in theoretical physics.

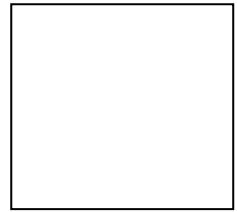

**Gerry Vernon Dozier**, Ph.D., is the Charles D. McCrary Eminent Chair Professor in the Department of Computer Science & Software Engineering at Auburn University. Dr. Dozier is the director of the Artificial Intelligence & Identity Research (AI2R) Lab. He is currently applying his research expertise in the areas of Artificial Intelligence, Machine Learning, Behavioral Analytics, and Evolutionary Computation to the areas of Cyber Identity Protection and Cyber Security. Dr. Dozier has published over 140 conference and journal publications. Dr. Dozier earned his Ph.D. in Computer Science from North Carolina State University.

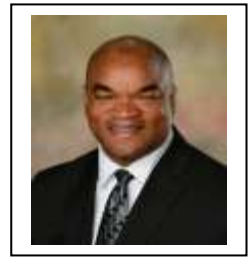